\documentstyle[12pt]{article}
\def \e1{\varepsilon}
\def\Z{Z \!\!\! Z}
\def\C{{\mathchoice {\setbox0=\hbox{$\displaystyle\rm C$}\hbox{\hbox
to0pt{\kern0.4\wd0\vrule height0.9\ht0\hss}\box0}}
{\setbox0=\hbox{$\textstyle\rm C$}\hbox{\hbox
to0pt{\kern0.4\wd0\vrule height0.9\ht0\hss}\box0}}
{\setbox0=\hbox{$\scriptstyle\rm C$}\hbox{\hbox
to0pt{\kern0.4\wd0\vrule height0.9\ht0\hss}\box0}}
{\setbox0=\hbox{$\scriptscriptstyle\rm C$}\hbox{\hbox
to0pt{\kern0.4\wd0\vrule height0.9\ht0\hss}\box0}}}}
\def\R{{\mathchoice {\setbox0=\hbox{$\displaystyle\rm R$}\hbox{\hbox
to0pt{\kern0.4\wd0\vrule height0.9\ht0\hss}\box0}}
{\setbox0=\hbox{$\textstyle\rm R$}\hbox{\hbox
to0pt{\kern0.4\wd0\vrule height0.9\ht0\hss}\box0}}
{\setbox0=\hbox{$\scriptstyle\rm R$}\hbox{\hbox
to0pt{\kern0.4\wd0\vrule height0.9\ht0\hss}\box0}}
{\setbox0=\hbox{$\scriptscriptstyle\rm R$}\hbox{\hbox
to0pt{\kern0.4\wd0\vrule height0.9\ht0\hss}\box0}}}}

\begin{document}
\begin{flushright} 
{hep-th/0102053}
\end{flushright}
\title{On root systems in spaces with degenerate metric}
\author{I.V.Kostyakov, N.A.Gromov, V.V.Kuratov \\
Department of Mathematics, \\
 Syktyvkar Branch of IMM UrD RAS,\\
e-mail: gromov@dm.komisc.ru }
\maketitle

\begin{abstract}
 A root systems in Carroll spaces with degenerate metric are defined.
It is shown that their Cartan matrices and reflection groups are affine.
With the help of the geometric consideration the root system structure
of affine algebras
is determined by a sufficiently simple algorithm.
\end{abstract}

It is known that nonsemisimple Lie groups and algebras may
be obtained by introducing the degenerate (nilpotent)
directions into the main notions of semisimple Lie groups
and algebras theory \cite{A1}.
It is algebraic approach to the limit passage from simple to
nonsemisimple groups suggested by E.In{\"o}n{\"u} and E.P.Wigner
\cite{A2}. A further kind of algebraic contractions is graded
contractions \cite{A3}.

All models of modern physics are appeared due to
underlying symmetries. Simple finite Lie algebras and
infinite affine Kac--Moody algebras takes an important part in
describing such symmetries.
Interest in models with nonsemisimple symmetry is growing
\cite{A4,A5} and this models may be investigated with the help
of contractions.
There are some models manufactured by contractions from some
initial ones, for example usual Toda chains may be obtained
from affine Toda chaines by contractions \cite{A6,A7},
in this case infinite-dimensional affine Lie algebra pass to
finite-dimensional simple Lie algebra.
The properties of models with different Lie algebras may be
related if there is the relations between their algebras.
In this paper relations between root systems of 
finite/infinite-dimensional algebras are proposed.
This relations are determined by geometric way. 

It is poorly known \cite{A1,2,3}, that all $3^n$ $n$-dimensional spaces
of constant curvature may be obtained by multiplications of an orthonormal
basis elements of Euclidean space by certain products of the parameters
$j_{k}=1,\iota_{k},i,\;  k=1, \ldots ,n,$
where the dual units $\iota_{k}$ are
nilpotent $\iota^2_k=0$ and obey the commutative law of multiplication
$ \iota_{k}\iota_{m}=\iota_{m}\iota_{k}\neq 0$.
The dual values of the parameters $j$ are corresponded to the
spaces with degenerate metrics. 
The main difference of affine root system is the presence
of a special (imaginary)  root $\delta$
with the nilpotent property $ \delta^{2}=0.$
A comparison of the nilpotent properties of the imaginary root
and of the dual unit suggests that root systems in
spaces with degenerated metric correspond to affine algebras.
This geometrical interpretation of root system of affine 
Lie algebras  may be a help in understanding their properties.

Kac-Moody algebras are characterized by Chevalley relations
\begin{equation}
[h_i,h_j]=0, \ \
[h_i,e_j]=a_{ij}e_j, \ \
[h_i,f_j]=-a_{ij}f_j, \ \
[e_i,f_j]=\delta_{ij}h_j,
\label{1}
\end{equation}
and  Serre relations
\begin{equation}
\left ( ad e_i \right )^{1-a_{ij}}e_j=
\left ( ad f_i \right )^{1-a_{ij}}f_j=0,
\label{2}
\end{equation}
where  $h_i, e_i, f_i, i=1,2, \dots ,r$ are Weyl generators.
Cartan matrix $ A=(a_{ij})$ may be given algebraically:
1) $a_{ij}\in \Z$, 2) $a_{ij}<0$ for $i \neq j$, 3) $a_{ij}a_{ji} \leq 3,$
4) $a_{ij}=0 \Longleftrightarrow a_{ji}=0$ ,
5) $det(a_{ij}) \neq 0 $ --- for a simple Lie algebras and
1), 2), 4), as well $3^a) $ $a_{ij}a_{ji} \leq 4,$ $\ 5^a)$
det$(a_{ij})=0,$
 the equation $\sum a_{ij}x_j=0$ has only one solution ---
for an affine algebras.
All such matrices are classified and hence
all simple and affine Lie algebras are listed.
Affine Kac--Moody algebras was discovered by algebraic way \cite{1}.

Cartan matrix may be also given  geometrically
with the help of a root system. A root system of a simple Lie algebra
is a finite set of vectors
$\tilde{\Pi}_0=\{\alpha,\beta,\dots \}$
in Euclidean space
$ V(r) $  with a positive defined nondegenerate metric if
1) $s_{\alpha}\tilde\Pi_0=\tilde\Pi_0$ for all
$\alpha \in \tilde\Pi_0;$
2)  for all $ \alpha, \beta \in \tilde{\Pi}_0, \
2n(\alpha,\beta)={(\alpha,\beta)\over (\alpha,\alpha)}
\in \Z$ are integers;
3) the linear span of  $\tilde{\Pi}_0$ coincides with $V(r):$
${\cal L}(\tilde{\Pi}_0)=V(r),$ where
\begin{equation}
s_{\alpha}(\beta)=\beta -2{(\alpha,\beta) \over (\alpha,\alpha)}
\alpha
\label{4}
\end{equation}
is the  reflection of $\beta$ with respect to $\alpha$.
A subset $K \subset \Pi_0$ is called a {\it basis}
$K=\{\alpha_i, i=1,\ldots ,r \}$ {\it of root system} $\Pi_0$
if the following conditions hold:
1) $K$ form a basis of $V(r),$
2) $\forall \alpha \in \Pi_0 \; \alpha=\sum_{i=1}^{r} m_i
\alpha_i,$ where $\alpha_i \in K, \ m_i \in \Z$
and all $m_i \geq 0$ (positive roots)
or all $m_i \leq 0$ (negative roots).
A matrix
\begin{equation}
(\tilde A)_{ij}=n(\alpha_i,\alpha_j)=
2{(\alpha_i,\alpha_j)\over (\alpha_i,\alpha_i)},\; \;
\alpha_i, \alpha_j \in K
\label{5}
\end{equation}
is called {\it a Cartan matrix} $\tilde A$ {\it of root system}.
A set of reflections form a group $W_0$ called a {\it Weyl group} of root
system.

Consider $(r+1)$-dimensional vector space $V(r,1)$ equipped with a
degenerate metric of signature $(0,+,\dots,\ +,),$ i.e.
the scalar product of vectors $\nu=(\nu_0,\nu_1,\dots,\nu_r)$ and
$w=(w_0,w_1,\dots,w_r)$ is define as $(\nu,w)=\sum_{k=1}^{r}\nu_k w_k.$
According with  geometric viewpoint $V(r,1)$ is a product bundle with
$r$-dimensional Euclidean subspace $V(r)$ as a base and  one-dimensional
orthogonal to $V(r)$ subspace as a fiber.  The projection is given by
mapping $\pi:V(r,1)\rightarrow V(r)$. At $r=3$ the space
$V(3,1)$ may be interpreted as one of the possible four-dimensional
space-time models, namely Carroll kinematic \cite{3}--\cite{5}.
Therefore we shall call $V(r,1)$ as   Carroll space.

 Above mentioned  property of the imaginary root gives reazons for the
following definitions. A vector set $ \tilde{\Pi} $ in Carroll space 
$V(r,1)$ is called {\it degenerate root system} (DRS), if  the properties
1)--3) of root system (with the replacement $ \tilde{\Pi}_0 $ by
$ \tilde{\Pi},\; $ $ V(r) $ by $ V(r,1) $) and
4) $ \forall \alpha \in \tilde{\Pi} \; \pi \alpha \neq 0 $
are satisfied.
A root system is called {\it reduced}
if there are no parallel roots (the roots $ \alpha $
and $ -\alpha $ are antiparallel).
A root system $\tilde{\Pi}$ is called {\it reducible}
if $V(r,1)=V_1\oplus V_2$ and $\tilde{\Pi}=\Pi_{1}\cup\Pi_{2},$ where
$\Pi_{1}$ and $\Pi_{2}$ are the root systems in
$ V_1 $ and $ V_2$ respectively. Otherwise a root system is called
{\it irreducible}.
The reduced irreducible root system will be denoted by $\Pi$.
A basis  $ B=\{ \alpha_i, i=0, \ldots ,r \}$ of $ \Pi$
is defined similarly to the ordinary root system.
 A matrix $ \bar{A},$ defined by (\ref{5}) for
$ \alpha_i, \alpha_j \in B $ is called {\it a Cartan matrix}
of DRS.
It is natural to associate with
the imaginary root the
vector $\varepsilon_0 $ in $V(r,1),$
pointing out the degeneracy  direction.

{\bf Proposition 1.}
{\bf \it The projection $\pi$
of degenerate root system $ \Pi $ onto $ V(r) $
gives the root system $ \tilde{\Pi}_{0}, \; $
$ \pi : \Pi \rightarrow \tilde{\Pi}_0. $
}

{\it Proof.}  Let us separate the vectors of DRS
$\Pi=\{\alpha, \beta, \ldots \}$ onto collinear and perpendicular to
$\varepsilon_0 $ components: $\alpha=\alpha_0+\alpha_{\bot}, $
$ \beta=\beta_0+\beta_{\bot}, \ldots.$
From the property 4) of DRS we get
$ \alpha_{\bot}\neq{0},\; \beta_{\bot}\neq{0}, \ldots .$
  The reflections in $\Pi$
 are passed into reflections in $ \tilde{\Pi}_0 $
 under  the projection $\pi:$
$ s_{\alpha_0+\alpha_{\bot}} \left ( \beta_0+\beta_{\bot} \right )=
s_{\alpha_{\bot}} \left ( \beta_{\bot} \right ),$
i.e. for $ \tilde{\Pi}_{0}=\{\alpha_{\bot}, \beta_{\bot}, \ldots \} $
 the property 1) of root system is held. Further from
$ n(\alpha,\beta)= n(\alpha_{\bot},\beta_{\bot}) $
it follows that the property 2) of root system is fulfilled.
Finally, from ${\cal L}(\Pi)=V(r,1)$ we obtain that
the linear span of $ \tilde{\Pi}_0 $ coincides with $ V(r), $
i.e. the property 3) is valid. Therefore $ \tilde{\Pi}_0 $ is
a root system of simple Lie algebra.

{\bf Remark.} An unreduced root system may be obtained at
the projection of the reduced DRS. Running ahead note
that it will occur if DRS
corresponds to the twisted affine Lie algebra  $\hat{A}_{2l}^{(2)}.$

{\bf Proposition 2.}  {\bf \it
The Cartan matrix of the degenerate root system is affine.}

{\it Proof.}
It is enough to show that the conditions
1), 2), $3^a$), 4), $5^a$) for affine Cartan matrix follow
 from the definition of degenerate root system.
The integerness 1) follows from the property 2) of DRS.
Further, similarly to the case of ordinary root system \cite{6} one may show
that the difference $\alpha-\beta$ is the root if $n(\alpha,\beta)>0,$
i.e. the angle between projections of vectors onto $V(r)$ is acute.
For simple (basis) roots $n(\alpha,\beta)<0,$
i.e. the angle between projections is obtuse
therefore the condition 2) is held.
Let $\alpha,\beta \in \Pi $ are an arbitrary roots.
We have the equation $n(\alpha,\beta)n(\beta,\alpha)=
4\cos \phi,$ where in contrast to nondegenerate case, $\phi$ is
the angle between the projections of vectors onto $ V(r).$
Since the collinearity of projections does not means the collinearity
of vectors itself, the case $\phi=0$
can not be rejected now, in full correspondence with
$3^{a}).$ The condition 4) follows from the symmetry of scalar product.
It remains to proof the conditions 5).
It is clear that all metric properties of DRS are determined by the
properties
of vector projections.  It was shown in Proposition 1 that these projections
form the root system of simple Lie algebra. All basis roots of DRS are
projected onto subspace $ V(r), $  forming nonacute angles
with each other. In this case one of the roots becomes linearly dependent
since the number of basis roots of root system,
forming by the projections of roots of DRS,
is just one less.
Therefore, the condition 5) is held for the Gramm matrix,
composed by the basis roots of DRS.
 Dividing its columns
by corresponding square lenght
 of basis roots, we obtain the same property for Cartan matrix.

Degenerated root systems are called {\it equivalent,}
if their Cartan matrices are the same. Since  Cartan matrix of DRS
is determined by its projection an equivalence class is formed by all DRS
with the same projections onto $ V(r). $
In  general case the projection $\Pi$ onto $V(r)$ is not contained in
$\Pi.$  $ \Pi $ is called  {\it canonical} DRS if
$ \pi\Pi \backslash P = \Pi_{0} \subset \Pi, $ where
$ P = \{  \beta \in \pi\Pi \; | \; 2\beta \in \pi\Pi \} $ or
$ P = \{  \beta \in \pi\Pi \; |\; \frac{1}{2}\beta \in \pi\Pi \},$
i.e. if  reduced root system $\Pi_{0}$ is contained in subspace $V(r)$.

Let us construct the basis
$ B =\{\alpha_{0},K\} = \{ \alpha_{0},\alpha_{1}, \ldots , \alpha_{r} \} $
of canonical DRS. It is clear that as $K$ should be taken the basis of
root system $\Pi_{0}.$
It remains to construct the $(r+1)$-th basis root
$ \alpha_{0} = \pi\alpha_{0} + \delta\epsilon_{0} \in V(r,1),
\;\delta \in R,$ that is in fact to
choose the projection $\pi\alpha_{0}$ in subspace
$V(r)$ in such a way  that the set of vectors $\{\pi\alpha_{0},K\}$
generate   according to  the Proposition 2 affine Cartan matrix.
It is clear that $\pi \alpha_0$ is such linear combination of vectors from
$K,$ that
$(\pi\alpha_{0}, \alpha_{k}) \leq 0, k=1, \ldots,r.$
As it was mentioned in Proposition 1
$ \tilde{\Pi}_{0} =\pi\Pi $ may be both reduced and
irreducible root system.

Let $ \tilde{\Pi}_{0}=\Pi_{0} $ be a reduced root system,
then $ \pi\alpha_{0}\in\Pi_{0}. $
When  roots of different length are presented
in $ K=\{\alpha_{1}, \ldots,\alpha_{r}\}, $ then two cases are
possible.

1) $ |\pi\alpha_{0}|=|\alpha_{l}|.$
All root systems $\Pi_0$ of simple Lie algebras have the highest vector
$\theta =  \sum_{k=1}^{r}a_{k}\alpha_{k},$ where
$ a_{k} $ are the marks on Dynkin diagramms of simple Lie algebras.
 The properties $ (\theta,\alpha_{k}) \leq 0, k=1, \ldots,r,$
are satisfied for this highest root
and also $|\theta|=|\alpha_{l}|$.
 (When all roots in $K$ has the same length,
we consider their as long roots).
By setting $\pi\alpha_{0} = -\theta$
we obtain the bases $B$ of root systems of all untwisted
affine algebras  \cite{1}.

 2) $ |\pi\alpha_{0}|=|\alpha_{s}|. $
Root systems $\Pi_{0}$ of simple Lie algebras
 $ C_r, B_r, F_4, G_2$ contain, except $\theta,$ the root
$\tilde{\theta}= \sum_{k=1}^{r}\tilde{a}_{k}\alpha_{k},$
where $ \tilde{a}_{k} $ are Kac marks on Dynkin diagramm of
affine algebras.
The properties
$(\tilde{\theta},\alpha_{k}) \leq 0, k=1, \ldots,r,$
are satisfied  and
 $|\tilde{\theta}|=|\alpha_{s}|.$
We obtain the bases $B$ of root systems of all twisted affine
Lie algebras, except $\hat{A}^{(2)}_{2r},$  by setting
$\pi\alpha_{0} = - \tilde{\theta}$  \cite{1}.

Let now $\tilde{\Pi}_{0} = \Pi_{0} \cup P$ be a nonreduced root
system (there are collinear vectors with the length twice as little
 of the length of long roots $\alpha_{l}\in\Pi_{0}$
or  with the length twice as much
 of the length of short roots $ \alpha_{s}\in\Pi_{0}).$
If $\pi\alpha_{0} \in \Pi_{0},$
then it follows from  above consideration,
that $\pi\Pi=\Pi_{0}$ is reduced root system.
But in our case $\tilde{\Pi}_{0}$ is nonreduced, therefore
the required DRS  can not be
 constructed  in this way. It remains to regard the case
$\pi\alpha_{0} \not\in \Pi_{0}$.
When $|\pi\alpha_{0}| =
\frac{1}{2}|\alpha_{l}|,$ then $\pi\alpha_{0} = -\frac{1}{2}\theta$
and when $|\pi\alpha_{0}| = 2|\alpha_{l}|,$ then
$ \pi\alpha_{0} = -2\theta,$ where $\theta$ is a highest root of $\Pi_{0}$.
In this case we obtain the bases $B$ of root systems
of twisted affine algebras
 $\hat{A}^{(2)}_{2r}$  \cite{1}.

Thus we have constructed the bases of root systems of all affine
Lie algebras, relying on the properties of root systems of simple
Lie algebras and on the properties of DRS.

Reflections  of DRS form  Weyl group $W$.
This group is generated by the basic reflections
$s_{\alpha_0},s_{\alpha_1},\dots ,s_{\alpha_r}.$
 It is obvious that last $r$ basic
reflections generate the subgroup $W_0.$
Consider the $k$-th power of translation operator of root $x$
along $\varepsilon_0$, which is constructed from reflections
\begin{equation}
t_{\alpha}^{k}(x)\stackrel{df}{=}s_{k\delta \varepsilon_0-\alpha}
s_{\alpha}(x)=x-2{(\alpha,x) \over(\alpha,\alpha)}
k\delta \varepsilon_0,
\end{equation}
where $\alpha, \ k\delta \varepsilon_0-\alpha, x \in \Pi$,
 $ (\alpha,x) \neq 0.$
The set of translations generates a commutative translation subgroup $T.$
It is easily to show that Weyl group $W$ is a semidirect product of its
subgroup $W=W_0\triangleright T.$

Now we determine the structure of DRS. It is enough to take
into account the fact, that the   root length
and its component $\e1_{0}$
are not changed under reflections:
$w(\delta\e1_{0})=\delta\e1_{0}, w \in W_{0},$
i.e.  subgroup $W_0$ acts transitively on the set of equal
length roots with the same $\e1_{0}$ coordinate and to construct
the shifts operators for different length roots.
The complete root  system is the join of imaginary and real roots:
$ \Pi=\Pi_{re} \cup \Pi_{im},$ where $\Pi_{im}=\{ n\delta\e1_{0}|n \in Z \}$.
It remains to build $\Pi_{re}$.

Consider first $\Pi_{re}$  of untwisted affine Lie algebras.
The root $\alpha_0$ has maximal length $|\alpha_0|=|\alpha_l|$ for
this algebras. The subset of shifted long roots
$\{ \Pi_{0}^{l} + \delta\e1_{0}\}$ is obtained by acting of reflections from
$W_0$ on $\alpha_{0}=\pi\alpha_0+\delta\e1_0, \; \pi\alpha_0 \in \Pi_0$.
The short root $\alpha_{m}$ is shifted by operator
$ t_{\alpha_{m-1}}(\alpha_{m})=\alpha_{m} + \delta\e1_{0},$
where $\alpha_{m-1}$ is a long root, connected with $\alpha_{m}$
on Dynkin diagramm. Here $m=1$ for $C_{r}^{(1)},\; m=2$ for
$G_{2}^{(1)},\; m=3$ for $F_{4}^{(1)},\; m=r$ for $B_{r}^{(1)}.$
Using  reflections from $W_{0}$ we obtain a  set
$\Pi_{0}^{s} + \delta\e1_{0}$ of shifted  short roots. As far as
$ \Pi_{0}^{l}\cup\Pi_{0}^{s}=\Pi_{0}$ for untwisted algebras, then
$\Pi_{0} \subset V(r)$ pass to $\Pi_{0}+\delta\e1_{0} \subset V(r,1).$
The next step is the shift of long root $\alpha_{0}: \ $
$t_{\alpha_{p}}(\alpha_{0}) = \alpha_{0} + \delta\varepsilon_0,$
where $p=6$ for $E_6^{(1)},\ p=2$ for $B_r^{(1)},D_r^{(1)},\ p=1$
for remaining untwisted affine algebras. The repetition of shift procedure
of long and short roots leads to the set $\{\Pi_0+2\delta\varepsilon_0\}$.
Shift of the set $\Pi_{0}$  in negative direction of axis
$\e1_{0}$ is constructed in a similar way.
 Thus DRS of untwisted affine Lie algebras is as follows
\begin{equation}
\Pi_{re}=\{\Pi_0+n\delta\varepsilon_0| \ n \in Z\}.
\label{10}
\end{equation}

For twisted algebras (except of $A_{2r}^{2}$)
the root $\alpha_0$ has minimal length $|\alpha_0|=|\alpha_s|$.
The subset of short
shifted roots $\{ \Pi_{0}^{s} + \delta\e1_{0}\}$
is obtained by acting of reflections from $W_0$ on
$\alpha_{0}=\pi\alpha_0+\delta\e1_0, \; \pi\alpha_0 \in \Pi_0$.
The long root $ \alpha_{m}$ is shifted by operator
$ t_{\alpha_{m-1}}(\alpha_{m})=\alpha_{m} +
2\delta\e1_{0},$ where $\alpha_{m-1}$ is a short nonorthogonal to $\alpha_m$
root. Here $m=1$ for $D_{r+1}^{(2)},\; m=2$ for $D_{4}^{(3)},\; m=3 $
for $ E_{6}^{(2)},\; m=r$  for $A_{2r-1}^{(2)}.$
The set $ \{\Pi_{0}^{l} + 2\delta\e1_{0}\}$
is obtained with the help of reflections from $W_{0}$.
The next step is the shift of short roots.
The root $\alpha_{0}$ is shifted by
$ t_{\alpha_{p}}(\alpha_{0}) = \alpha_{0} + \delta\varepsilon_0
= \pi\alpha_0+2\delta\e1_0, \;$
 $p=1$ for $D_{4}^{(3)}, E_6^{(2)}, \ p=2$
for $A_{2r-1}^{(2)}$ and the root $\alpha_r$ is shifted by
$t^2_{\alpha_{r-1}}(\alpha_r)=\alpha_r+2\delta\e1_0$
in the case of $D_{2r-1}^{(2)}$ algebra.
(The operator $t_{\alpha_{r-1}}=s_{\delta\e1_0-\alpha_{r-1}}s_{\alpha_{r-1}}$
is not defined now because $\delta\e1_0-\alpha_{r-1} \not \in \Pi,$
therefore we have use the operator $ t^{2}_{\alpha_{r-1}})$.
The shifted short roots are  transformed by subgroup $W_{0}$
to the set $\{ \Pi_{0}^{s} + 2\delta\e1_{0}\}$.
  The set $\{\Pi_{0}^{l}+2\delta\e1_{0}\}$ of  shifted long roots
  is transformed by $W_0$ and by
the   operator $t^{2}_{\alpha_{m-1}}(\alpha_{m})=\alpha_{m} +
4\delta\e1_{0}$
to the set $\{\Pi_{0}^{l}+4\delta\e1_{0}\}.$
$\Pi_{re}$ of twisted affine algebras (except of $A^{(2)}_{2r}$)
is the join of the sets of  long and short shifted roots.
\begin{equation}
\Pi_{re}=\{\Pi_0^s+n\delta\varepsilon_0\} \cup
 \{\Pi_0^l+nk\delta\varepsilon_0\}, \ n \in Z,
\ k=2,3.
\label{11}
\end{equation}

Now we find the construction of $\Pi_{re}$ for algebras $A_{2r}^{(2)},$
where $|\alpha_0|={1\over2}|\alpha_l| <|\alpha_s| <|\alpha_l|$
and $\pi\alpha_0 \not \in \Pi_0$. The operator $t_{\alpha_0}$
shifts the short root $\alpha_1 \in \Pi_0, \ |\alpha_1|=
|\alpha_s|$ on $2\delta\varepsilon_0$:
$t_{\alpha_0}(\alpha_1)=\alpha_1+2\delta\varepsilon_0.$ The operator
$t_{\alpha_1}=s_{(\delta\varepsilon_0-\alpha_1)}s_{\alpha_1}$ is not defined
since $\delta\varepsilon_0-\alpha_1 \not \in \Pi,$
therefore the shift operators of short roots $\alpha_k$
look as follows:
$t^{2}_{\alpha_{k}}(\alpha_{k+1})=\alpha_{k+1} + 2\delta\e1_{0},
\; k=1,2,\ldots,r-2.$
  The shifted short roots
are transformed by $W_0$
to the set $ \Pi_0^s + 2\delta\varepsilon_0.$ The short roots are shifted
with the help of operators $t_{\alpha_0}^n,$ $t_{\alpha_k}^n, \ n \in Z$
on $2n\delta\varepsilon_0$ and then are reproduced by $W_0$.
As a result we obtain a subset
$\Pi_1=\{\Pi_0^s+2n\delta\varepsilon_0| n \in Z \}.$
The long root $\alpha_r \in \Pi_0, \ |\alpha_r|=|\alpha_l|$
is shifted on $4\delta\varepsilon_0:$
$ t^2_{\alpha_{r-1}}(\alpha_r)=\alpha_{r}+4\delta\varepsilon_0.$
Similarly, using the shift operators $t_{\alpha_{r-1}}^{2n}$
and reflections from $W_0,$ we obtain the subset of long roots
$\Pi_2=\{\Pi_0^l+4n\delta\varepsilon_0 | n \in Z \}$.

It remains to explain the construction of roots of length
$ |\alpha_0|={1\over2}|\alpha_l|.$ The reflections from $W_0$
transform the basis root $\alpha_0 = \pi\alpha_0 + \delta\varepsilon_0$
to the set of roots $\{ {1\over2} \Pi_0^l+\delta\varepsilon_0 \},$
since $\pi \alpha_0 = -{1\over2} \theta$.
Operator $t_{\alpha_1}^2$ shift $\alpha_0$ on $2\delta\varepsilon_0$:
$t_{\alpha_1}^2(\alpha_0)=\alpha_0 +2\delta\varepsilon_0
= \pi\alpha_0+3\delta\varepsilon_0$ and then this shifted root
is reproduced by reflections
from subgroup $W_0$  to the set
$\{ {1\over2}\Pi_0^l+3\delta\varepsilon_0 \}.$
Operators  $t_{\alpha_1}^{2n}, \ n \in Z$ and reflections from
$W_0$ generate the subset $ \Pi_3=\{{1\over2}\Pi_0^l+
(2n+1)\delta\varepsilon_0 | n \in Z \}.$ Then the set of real roots of
$A_{2r}^{(2)}$ is the join of subsets
\begin{equation}
\Pi_{re}=\{\frac{1}{2} \Pi_0^l+(2n+1)\delta\varepsilon_0\} \cup
\{\Pi_0^s+2n\delta\varepsilon_0\}
\cup \{\Pi_0^l+4n\delta\varepsilon_0\}, \ n \in Z.
\end{equation}

Thus, whith the help of geometric consideration
using shifts operators of basic roots
along $\varepsilon_0$ and reflections from subgroup $W_0$
we have obtained sufficiently simply the structure of DRS
in Carroll spaces.
It is natural that
our results are coincided with the analogous ones
for affine root systems in pure algebraic consideration.

An analysis of other types of Kac-Moody
algebras which are discussed in literature
shows that their Cartan matrices may be obtained with the help of
geometric consideration.
For example, root systems in pseudoeuclidean spaces are connected
with hyperbolic Kac-Moody algebras \cite {7} and
Borcherds algebras \cite{8},\cite{9}. In this case the algebraic
condition $a_{ii}=0$ on symmetrized Cartan matrix  is corresponded
geometrically to the case, when a basic root belong to the con $(x,x)=0$
and the condition $a_{ii}<0$ means, that  basis roots have both
positive and negative length, i.e. they are situated in different
con sectors.

Root systems in the space with twice  degenerate metric are connected
apparently with toroidal algebras \cite{10},\cite{11}.
This geometric interpretation make possible to establish
some properties of their Cartan matrices. So the condition $5^a$)
is changed and the property $ a_{ij}<0$ at $i \neq j$
is not more held
since it is impossible to distribute
$n+2$ projections of basis roots on $n$-dimensional Euclidean space
so that all angles between these roots are nonacute,
i.e. all scalar products between them are nonpositive.
It is clear that this imply the modification of Serre relations.

A generalization of the observation indicated in the letter
 leads to the suggestion:
{\it
to regard root systems
in spaces of constant curvature (especially with degenerate metric)
 and to construct  corresponding Lie algebras.
}
Work on this problem is continued. 

\end{document}